\documentclass[conference]{IEEEconf}

\input epsf
\usepackage{caption,subcaption}
\usepackage{listings}
\usepackage{xcolor}
\usepackage{enumitem}
\usepackage{fancyvrb}
\usepackage{multirow}
\usepackage{graphicx}
\usepackage{adjustbox}
\usepackage{tcolorbox}
\usepackage{comment}
\usepackage{refcount}
\usepackage{url} 
\usepackage{amsmath}
\usepackage{booktabs}
\usepackage{hyperref}
\usepackage{cleveref}
\usepackage{balance}

\definecolor{superlightgray}{HTML}{EEEEEE}
\definecolor{lightblue}{HTML}{DDDDFF}
\definecolor{lightorange}{HTML}{FFD580}
\definecolor{darkorange}{HTML}{EB8034}
\definecolor{darkgreen}{HTML}{005500}

\newcommand{\snippettransformer}{CodeCocoon }

\usepackage{color}
\usepackage{soul}
\definecolor{highlight}{rgb}{0.9,1.0,0.9}
\definecolor{hl}{rgb}{0.80,0.80,0.80}
\sethlcolor{hl}

\begin{document}

\title{A Metamorphic Testing Approach to Diagnosing Memorization in LLM-Based Program Repair}

\author{Milan De Koning$^{1}$, Ali Asgari$^{2,*}$, Pouria Derakhshanfar,$^{1}$, Annibale Panichella$^{2,*}$\\
	\normalsize $^{1}$JetBrains Research, Amsterdam, The Netherlands\\
	\normalsize $^{2}$Delft University of Technology, Delft, The Netherlands\\
	\normalsize milan.de.koning@jetbrains.com, A.Asgari-2@tudelft.nl, pouria.derakhshanfar@jetbrains.com, A.Panichella@tudelft.nl\\
	\normalsize *corresponding author
}

\maketitle

\begin{abstract}
LLM-based automated program repair (APR) techniques have shown promising results in reducing debugging costs. However, prior results can be affected by data leakage: large language models (LLMs) may memorize bug fixes when evaluation benchmarks overlap with their pretraining data, leading to inflated performance estimates. 
In this paper, we investigate whether we can better reveal data leakage by combining metamorphic testing (MT) with negative log-likelihood (NLL), which has been used in prior work as a proxy for memorization. We construct variant benchmarks by applying semantics-preserving transformations to two widely used datasets, Defects4J and GitBug-Java. Using these benchmarks, we evaluate the repair success rates of seven LLMs on both original and transformed versions, and analyze the relationship between performance degradation and NLL.
Our results show that all evaluated state-of-the-art LLMs exhibit substantial drops in patch generation success rates on transformed benchmarks, ranging from $-$4.1\% for \textit{GPT-4o} to $-$15.98\% for \textit{Llama-3.1}. Furthermore, we find that this degradation strongly correlates with NLL on the original benchmarks, suggesting that models perform better on instances they are more likely to have memorized. These findings show that combining MT with NLL provides stronger and more reliable evidence of data leakage, while metamorphic testing alone can help mitigate its effects in LLM-based APR evaluations.
\end{abstract}
\vspace{1.5ex}

\noindent\textit{Keywords—}Automated Program Repair, Metamorphic Testing, Large Language Models, Data Leakage,  AI for SE

\section{Introduction}
\label{sec:intro}

Debugging is a time-consuming activity in software development. Developers report spending 20–-60\% of their time on debugging tasks~\cite{beller_dichotomy_2018}, and software failures were estimated to cost \$1.56~trillion in the US alone in 2020~\cite{krasner2021cost}. Automating even a portion of this effort could lead to substantial economic and productivity gains. Thus, a growing body of research has focused on \textit{Automated Program Repair} (APR)~\cite{huang_survey_2023}, with recent advancements using \textit{Large Language Models} (LLMs) demonstrating remarkable effectiveness across many coding tasks, including program repair~\cite{zhang_systematic_2024, xiang_how_2024, xia_automated_2024}.

Recent progress in LLM-based program repair has focused on improving patch generation success rates on established benchmarks~\cite{xia_automated_2024, xiang_how_2024, li_hybrid_2024}, such as Defects4J~\cite{just_defects4j_2014}. However, emerging evidence suggests that these results may be inflated due to \textit{data leakage}\cite{zhou_dont_2023, sallou_breaking_2024, ramos_are_2024}, where evaluation data overlaps with pretraining corpora, allowing LLMs to memorize bug-fix pairs rather than generalize. For example, Ramos et al.\cite{ramos_are_2024} show that widely used APR benchmarks likely overlap with the training data of many open-source LLMs.

To mitigate data leakage, several studies propose benchmarks built from code committed after model pretraining cutoffs~\cite{wu_condefects_2024, zhang_critical_2024, zhou_lessleak-bench_2025}. However, the rapid release cycle of LLMs and the heterogeneity of their training data make leakage difficult to eliminate. Even with time-filtered datasets, overlap may persist, as bug-fixing commits and test cases are often duplicated across blogs, academic papers, and public repositories that are commonly included in pretraining corpora, especially for general-purpose models such as \texttt{ChatGPT}. Consequently, leakage remains hard to detect and prevent, even with carefully curated benchmarks. Moreover, maintaining such datasets requires substantial manual effort and they may still become outdated as models evolve. 

In this paper, we present a systematic empirical study of whether metamorphic testing (MT) can be used to analyze and potentially address data leakage in LLM-based APR. In our setting, metamorphic transformations modify the syntactic form of faulty code while preserving its semantics and behavior~\cite{applis_assessing_2021}. Examples include renaming identifiers, reordering statements, or rewriting loops~\cite{applis_assessing_2021, segura_survey_2016}. We then examine whether performance degradation on transformed code is associated with memorization rather than semantic generalization.

To guide our study, we formulated and investigated the following research questions:
\begin{itemize}
    \item \textbf{RQ1:} \textit{What is the impact of metamorphic transformations on the performance of LLM-based program repair?}
    \item \textbf{RQ2:} \textit{How is the observed LLMs' performance drop under metamorphic transformations related to potential data leakage in the benchmark code?}
    \item \textbf{RQ3:} \textit{Which types of code patterns and metamorphic transformations are most indicative of memorization effects in LLM-based program repair?}
\end{itemize}

Together, these questions position MT not only as a robustness testing technique but also as a diagnostic tool for identifying performance inflation due to benchmark contamination. To answer them, we study two benchmarks, Defects4J~\cite{just_defects4j_2014} and GitBug-Java~\cite{silva_gitbug-java_2024}. Defects4J is a widely used benchmark for LLM-based APR, but its bugs and fixes predate the cutoff dates of the evaluated LLMs and are therefore more likely to be affected by data leakage. In contrast, GitBug-Java contains more recent defects that are post cutoff dates, making it less likely to overlap with training data~\cite{silva_gitbug-java_2024}.
We apply a set of semantics-preserving transformations to both datasets that reflect natural coding variations~\cite{kim2016automatic}, such as identifier renaming with synonyms~\cite{panichella2025metamorphic}, as well as control-flow and formatting modifications~\cite{asgarimetamorphic}.

We evaluate seven state-of-the-art LLMs on both original and transformed benchmarks: ChatGPT-4o\footnote{\label{fn:openai}OpenAI, \url{https://openai.com}}, ChatGPT-4o-mini\footnotemark[\getrefnumber{fn:openai}], Claude-3.7-Sonnet\footnote{\label{fn:anthropic}\url{https://www.anthropic.com}}, Llama 3.1 8B~\cite{grattafiori_llama_2024}, Gemma 2 27B~\cite{team_gemma_2024}, Mistral 7B v0.3~\cite{jiang_mistral_2023}, and StarCoder 2 7B~\cite{lozhkov_starcoder_2024}.
All models show reduced success rates on transformed code, with average drops between 4.1\% and 15.98\% on Defects4J. GitBug-Java exhibits smaller effects, but still shows substantial drops on individual bugs. We find that these declines correlate with models’ negative log-likelihood (NLL) on the original benchmarks~\cite{ramos_are_2024}, supporting its use as a potential indicator of bug/fix memorization. 
Thus, combining MT with NLL-based memorization can provide stronger evidence of data leakage.

Our contributions are as follows:
\begin{itemize}
    \item We present a large-scale empirical study examining how metamorphic testing outcomes relate to potential data leakage in LLM-based program repair.
    \item We introduce \textbf{\snippettransformer}\hspace{-3pt}, a transformation framework for constructing realistic, semantics-preserving benchmark variants, which can be extended to other benchmarks and code-related tasks.
    \item We evaluate two APR benchmarks and seven state-of-the-art LLMs, showing consistent performance degradation under metamorphic transformations and complementary evidence of data leakage when considered alongside memorization signals.
    \item We argue that selectively evaluating metamorphically transformed variants as part of APR evaluation pipelines can help, together with NLL, diagnose performance inflation caused by benchmark contamination.
\end{itemize}



\section{Background and Related Work}
\label{sec:background}

\textbf{Automated Program Repair}.
Automated Program Repair (APR) aims to automatically generate patches for software bugs~\cite{durieux_empirical_2019} and it is typically structured in three steps: (1) fault localization, (2) patch generation, and (3) patch validation.

\textit{1) Fault Localization}.
This step identifies potentially faulty code locations that should be modified. APR tools often rely on existing fault localization (FL) techniques~\cite{wong_survey_2016}. Although accurate localization is critical for effective repair, many APR studies assume perfect fault localization and focus instead on patch generation and validation~\cite{huang_survey_2023}.



\textit{2) Patch Generation}.
Given a localized defect, APR tools generate candidate patches using a variety of techniques. These can be broadly categorized as follows~\cite{huang_survey_2023}: \textit{search-based} approaches use metaheuristics to explore large repair spaces~\cite{le_goues_genprog_2012, sidiroglou-douskos_automatic_2015}; \textit{constraint-based} techniques encode repair as a satisfiability problem~\cite{wei_automated_2010, chen_contract-based_2021}; \textit{template-based} methods apply predefined fix patterns~\cite{kim_automatic_2013, le_history_2016-1, tan_relifix_2015}; and \textit{learning-based} approaches train neural models to learn repair transformations~\cite{tufano_empirical_2019, chen_sequencer_2021}.
%
%
Among learning-based methods, those based on large language models (LLMs) have recently achieved state-of-the-art results~\cite{zhang_systematic_2024, xia_automated_2024, xiang_how_2024}. LLM-based APR can be realized through different strategies~\cite{zhang_systematic_2024}: \textit{fine-tuning} task-specific models such as CodeT5~\cite{wang_codet5_2021} and CodeBERT~\cite{feng_codebert_2020}; \textit{few-shot learning} with in-context examples, typically applied to medium-sized models like CodeX~\cite{chen_evaluating_2021}; and \textit{zero-shot prompting}, where general-purpose models such as ChatGPT generate patches without fine-tuning~\cite{zhang_systematic_2024}.

\textit{3) Patch Validation}.
Once candidate patches are generated, they must be assessed for correctness. State-of-the-art tools often generate hundreds of candidates per bug; for instance, SRepair~\cite{xiang_how_2024} produces up to 500 patches per defect. The most widely adopted validation approach is \textit{test-suite-based repair}, where a patch is considered valid if it compiles and passes all test cases~\cite{durieux_empirical_2019, xiang_how_2024, xia_automated_2024}. However, passing the test suite (or \textit{test-adequate}) does not necessarily imply correctness. Many patches exploit deficiencies in the test suite, such as by deleting buggy code or hard-coding values, resulting in overfitting. These patches may appear correct but fail to generalize. 

To distinguish between different outcomes, Durieux et al.~\cite{durieux_empirical_2019} proposed a taxonomy of patch validity: \textit{uncompilable} patches fail to compile; \textit{failing} patches compile but do not pass all tests; \textit{plausible} patches pass the test suite but may be incorrect; and \textit{correct} patches are semantically equivalent to the developer fix. Correctness, in this context, is typically verified via manual inspection or oracle comparison, as test suites alone are not sufficient.

\smallskip

\textbf{Large Language Models and Data Leakage}.
Large Language Models (LLMs) have demonstrated strong performance across a wide range of coding tasks, including code generation~\cite{jiang_survey_2024}, code summarization~\cite{sun_source_2024}, vulnerability detection~\cite{zhou_large_2024}, and automated program repair (APR)~\cite{zhang_systematic_2024}. Many of these models are fine-tuned to optimize specific objectives~\cite{naveed_comprehensive_2024}, enabling them to generalize across diverse tasks with minimal task-specific data. 

While LLMs often perform well on standard benchmarks, recent studies show that their robustness to simple, semantics-preserving code transformations remains limited~\cite{applis_assessing_2021, applis_searching_2023, zhou_evolutionary_2024, gao_discrete_2023, zhou_adversarial_2022, jia_clawsat_2023}. These findings raise concerns about whether high benchmark scores reflect genuine semantic understanding or surface-level pattern matching.

A related and more insidious issue is \textit{data leakage}, which occurs when evaluation benchmarks overlap with an LLM’s pretraining data, leading to inflated performance metrics and poor generalization~\cite{lopez_inter-dataset_2025, zhou_dont_2023}. This is particularly concerning in APR, where widely used benchmarks such as Defects4J~\cite{just_defects4j_2014} are likely to be present in pretraining corpora. As a result, models may generate correct patches by recalling memorized bug-fix patterns rather than reasoning over novel bugs.

Prior work has shown that LLMs can memorize and reproduce both natural language and source code, with larger models exhibiting stronger memorization effects~\cite{al-kaswan_traces_2024}. As a result, several studies in the machine learning and software engineering communities have investigated indirect signals for detecting such memorization in evaluation benchmarks.
A widely used approach relies on \emph{perplexity} and \emph{negative log-likelihood} (NLL), which measure how surprising a token sequence is to a model~\cite{xu_benchmarking_2024, li_estimating_2023}. NLL is defined as:
\begin{equation}
\small
NLL(x_i) = -\log p_\theta(x_i \mid x_{<i}),
\end{equation}
where $p_\theta$ denotes the model’s next-token probability distribution. Lower NLL (and equivalently lower perplexity) indicates that a sequence appears highly familiar to the model, which is consistent with prior exposure during training. Consequently, NLL has become a \emph{standard proxy} for model familiarity and memorization in large-scale LLM evaluations~\cite{xu_benchmarking_2024, li_estimating_2023}.

Ramos et al.~\cite{ramos_are_2024} applied NLL-based analysis to Java benchmarks such as Defects4J and GitBug-Java~\cite{silva_gitbug-java_2024}, comparing them against post-cutoff repositories. They found that several models exhibit substantially lower NLL on Defects4J, suggesting memorization of benchmark content and highlighting the risks associated with older benchmarks.
Importantly, while NLL is a well-established and widely accepted indicator of memorization~\cite{xu_benchmarking_2024, li_estimating_2023}, it does not, by itself, constitute definitive proof of data leakage. In this work, we therefore interpret NLL as a supporting signal and analyze it in conjunction with metamorphic testing outcomes, using converging evidence rather than relying on  single metrics in isolation.

\smallskip
\textbf{Data leakage mitigation}.
Several studies propose mitigating data leakage by creating new benchmarks~\cite{wu_condefects_2024, zhang_critical_2024, zhou_lessleak-bench_2025}, as also recommended by Zhang et al.~\cite{zhang_survey_2023}. However, benchmarks can quickly become obsolete as LLMs are continually retrained on newly scraped data. Bradbury and More~\cite{bradbury_addressing_2024} proposed to dynamically instantiate benchmark variants from handcrafted templates, showing consistent performance drops on HumanEval~\cite{chen_evaluating_2021}, albeit at substantial manual cost.

An alternative strategy is to apply \textit{metamorphic transformations} to existing benchmarks in order to reduce sensitivity to memorized instances while preserving their utility. Prior work has primarily applied metamorphic testing to study robustness and consistency of code models, without explicitly examining its relationship to benchmark contamination and memorization~\cite{asgarimetamorphic}. In this study, we explore metamorphic testing from a leakage-aware perspective, looking at how performance degradation under transformations relates to signals of model familiarity, such as NLL. By combining these signals with metamorphic transformations, we offer complementary evidence for identifying data leakage in LLM-based APR.

\smallskip
\textbf{Metamorphic testing}.
%
Metamorphic testing evaluates the robustness of models for code-related tasks by transforming the syntax or abstract syntax tree (AST) of input code without altering its semantics~\cite{applis_assessing_2021, applis_searching_2023}. By comparing model outputs on original and transformed inputs, one can assess robustness. This principle also applies to APR: if a model can fix a bug in the original snippet, it should also fix it in a semantically equivalent, transformed version. Prior work shows that many models are sensitive to irrelevant transformations, such as identifier renaming or structural changes, leading to inconsistent outputs~\cite{applis_assessing_2021, applis_searching_2023, zhou_evolutionary_2024, gao_discrete_2023, zhou_adversarial_2022, jia_clawsat_2023}. 
Natural transformations are crucial to avoid false alarms, as unnatural code may confuse models even though it would not occur in practice~\cite{yang_natural_2022, le-cong_evaluating_2024, zhou_evolutionary_2024}. Accordingly, in this study we apply several natural transformations, including identifier renaming with synonyms~\cite{panichella2025metamorphic}, as detailed in \Cref{sec:apr_setting}.

Several studies have applied metamorphic testing to deep code models, particularly in the context of APR~\cite{asgarimetamorphic}. Ge et al.~\cite{ge_robustnpr_2024} evaluated four neural APR models (Recoder, CoCoNut, SequenceR, and Tufano) and showed that even the most robust model failed on a substantial fraction (20\% ) of transformed cases. 
Le-Cong et al.~\cite{le-cong_evaluating_2024} studied neural APR models with an emphasis on transformation naturalness, finding that a significant portion of prediction changes were caused by unnatural transformations. They argued that such cases should be excluded and explored automated naturalness assessment using early LLMs based on cross-entropy. More recently, Li et al.~\cite{li_evaluating_2025} evaluated LLM-based repair on transformed Defects4J bugs and attributed observed performance drops to robustness issues, without considering data leakage. Xue et al.~\cite{xue_exploring_2024} evaluated open-source LLMs under metamorphic testing and proposed reverting transformations prior to repair, although the generalizability of this approach remains unclear.

Overall, prior work has established that code models are fragile under metamorphic transformations, but has largely focused on robustness analysis of neural APR systems or open-source LLMs. In contrast, our work evaluates state-of-the-art proprietary LLMs (e.g., ChatGPT-4o and Claude-3.7) and explicitly examines whether performance degradation under metamorphic transformations is associated with data leakage. We further provide a correlation analysis between metamorphic robustness and negative log-likelihood (NLL), offering a complementary, model-agnostic lens into LLM behavior beyond internal metrics alone~\cite{xu_benchmarking_2024, li_estimating_2023}.

\section{Methodology}
\label{sec:methodology}

\begin{figure}
    \centering
    \includegraphics[width=0.9\linewidth]{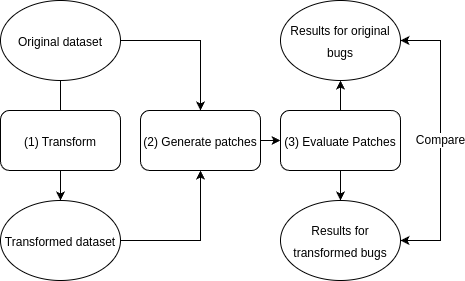}
    \caption{Experimental pipeline}
    \label{fig:pipeline}
\end{figure}

This section describes the methodology we applied to answer our three research questions:
\begin{itemize}
    \item \textbf{RQ1:} \textit{What is the impact of metamorphic transformations on the performance of LLM-based program repair?}
    \item \textbf{RQ2:} \textit{How is the observed LLMs' performance drop under metamorphic transformations related to potential data leakage in the benchmark code?}
    \item \textbf{RQ3:} \textit{Which types of code patterns and metamorphic transformations are most indicative of memorization effects in LLM-based program repair?}
\end{itemize}
To address these questions, we designed an experimental pipeline that is illustrated in \Cref{fig:pipeline}. The pipeline consists of three main stages. First, in the \textit{Transform} stage, we apply a suite of semantics-preserving code transformations—such as identifier renaming and control-flow rewriting—to introduce structural diversity while maintaining program behavior. Second, in the \textit{Generate Patch} stage, we use large language models (LLMs) to generate bug-fixing patches for both the original and transformed buggy code snippets. Each model is prompted in a Chain-of-Thought setting with the buggy function and its associated test cases. 
Finally, in the \textit{Evaluate Patch} stage, we validate generated patches by executing the project’s test suite. A patch is considered successful if it compiles and passes all tests. We compare success rates between original and transformed benchmarks to quantify robustness effects (RQ1), analyze how performance degradation correlates with memorization indicators such as negative log-likelihood (RQ2), and examine which transformation types most strongly expose memorization effects (RQ3).
All code and experimental artifacts are available at \url{https://zenodo.org/records/15837296}.

\textbf{Datasets.} \noindent We use the Defects4J~\cite{just_defects4j_2014} and GitBug-Java~\cite{silva_gitbug-java_2024} datasets. Defects4J contains 854 real-world bugs collected from mature Java projects spanning multiple years, with most bugs written in Java 1.6 to Java 1.8. GitBug-Java, by contrast, is a more recent benchmark containing 199 bugs collected between January 2023 and October 2023. It was explicitly designed to reduce the risk of data leakage by targeting newer defects that are less likely to appear in LLM pretraining corpora. Note that we excluded 44 bugs from GitBug-Java, as they do not involve single-function defects. For consistency across both benchmarks, our study focuses exclusively on function-level bugs, where the fix is applied to a single method.

\textbf{Transformations.} We developed \snippettransformer to support metamorphic transformations of code models. The framework applies a range of transformations to Java code snippets, including structural changes and identifier renaming, while ensuring the transformed code remains natural and human-readable. Although several alternative tools exist, \snippettransformer is, to the best of our knowledge, the only tool that systematically applies natural transformations and has been extensively validated to preserve program behavior. While our study focuses on APR, \snippettransformer is broadly applicable to other code-related tasks.
We implement a set of structural and renaming transformations, deliberately excluding unnatural ones, as they may introduce unrealistic inputs that reduce the validity of model evaluations~\cite{le-cong_evaluating_2024, yang_natural_2022}. Our transformations are selected from the set proposed in~\cite{le-cong_evaluating_2024}, where naturalness was assessed via human evaluation. We chose only those transformations rated highly for naturalness and applicable to typical Java code, ensuring realism without sacrificing diversity. \Cref{tab:chosen_transformations} lists the selected transformations. For variable and parameter renaming, we use ChatGPT-4o-mini\footnotemark[\getrefnumber{fn:openai}] to generate context-aware synonyms and acronyms, following best practices suggested in prior work~\cite{panichella2025metamorphic}.

\begin{table*}
\centering
\caption{The metamorphic transformations we apply in this study.}
\label{tab:chosen_transformations}
\small
\setlength{\tabcolsep}{4pt}
\begin{tabular}{l p{4cm} p{7.5cm}}
\toprule
\textbf{Transformation} & \textbf{Explanation} & \textbf{Example} \\
\midrule

Rename Function (RFun) & Replace the name of a function with a context synonym. &
\texttt{int add(int a, int b)} $\rightarrow$ \texttt{int sum(int a, int b)} \\

Rename Parameter (RPar) & Replace the name of a parameter with a context synonym. &
\texttt{void remove(int element)} $\rightarrow$ \texttt{void remove(int value)} \\

Rename Variable (RVar) & Replace the name of a local variable with a context synonym. &
\texttt{int total;} $\rightarrow$ \texttt{int result;} \\

For to While (F2W) & Replace for loop with while loop &
\texttt{for (int i = 0; i < n; i++)} $\rightarrow$ \texttt{int i = 0; while (i < n) \{ i++; \}} \\

Reverse If (RevIf) & Negate condition and swap if and else blocks &
\texttt{if(a)\{doA();\} else \{doB();\}} $\rightarrow$ \texttt{if(!a)\{doB();\} else \{doA();\}} \\

Nest Else If (NestEI) & Replace else-if with nested if inside else &
\texttt{if(a)\{\} else if(b)\{doB();\}} $\rightarrow$ \texttt{if(a)\{\} else \{ if(b)\{doB();\}\}} \\

Swap Equals operands (SEO) & Swap operands of equals expression &
\texttt{a == b} $\rightarrow$ \texttt{b == a} \\

Swap Relational operands (SRO) & Swap operands of relational operator &
\texttt{a > b} $\rightarrow$ \texttt{b < a} \\

Expand Unary increment (EUI) & Expand unary increment/decrement &
\texttt{i++} $\rightarrow$ \texttt{i += 1} \\

\bottomrule
\end{tabular}
\end{table*}
\textbf{Models.} \noindent We evaluate seven state-of-the-art models for automated code repair. This includes three closed-source models: ChatGPT-4o\footnotemark[\getrefnumber{fn:openai}], ChatGPT-4o-mini\footnotemark[\getrefnumber{fn:openai}], and Claude-3.7-Sonnet\footnotemark[\getrefnumber{fn:anthropic}]. In addition, we consider four open-source models: Llama 3.1 8B~\cite{grattafiori_llama_2024}, StarCoder2 7B~\cite{lozhkov_starcoder_2024}, Gemma 2 27B~\cite{team_gemma_2024}, and Mistral 7B v0.3~\cite{jiang_mistral_2023}.
Because our evaluation relies on instruction-following capabilities, we use instruction-tuned variants of the open-source models from Hugging Face\footnote{\label{fn:huggingface}\url{https://huggingface.co}}, and run them locally using vLLM~\cite{kwon2023efficient}. We exclude CodeGen 6B~\cite{nijkamp_codegen_2023} due to the absence of an instruction-tuned version on Hugging Face and lack of support in vLLM. We also omit Llama 70B~\cite{grattafiori_llama_2024} due to computational resource constraints.


\textbf{ APR settings.}
\label{sec:apr_setting} In our experiments, we apply APR by prompting the model under test similar to the SRepair method proposed by Xiang et al. \cite{xiang_how_2024}. We use Chain-of-Thought \cite{wei_chain--thought_2022} techniques to prompt our APR model to analyze the root cause of the bug, suggest solutions, and implement these solutions. In our prompt, we provide the buggy function, Javadoc, a random trigger test case, and the stack trace of this trigger test. 
We extract the patches from the response and replace the buggy function with the generated patches in the project before executing the tests. The key difference from SRepair \cite{xiang_how_2024} is that we let the LLM reason about the root cause and implement the solution simultaneously, reducing inference time and cost. This setup is sufficient to demonstrate robustness and data leakage issues in LLMs.
Finally, we assume function-level fault localization, meaning the model is given the entire buggy function but not the exact faulty line. This contrasts with prior studies~\cite{ge_robustnpr_2024, le-cong_evaluating_2024}, which assume perfect line-level fault localization. Function-level localization offers a more realistic and practical setting: in practice, identifying the faulty function requires significantly less manual effort or tooling precision than pinpointing the exact faulty line. Despite being coarser, this still provides enough context for LLMs to reason about and fix bugs, while better reflecting real-world usage scenarios.


\textbf{ Evaluating Robustness to Code Transformations}. 
To answer \textbf{RQ1}, we measure how robust LLMs are to semantics-preserving code transformations by comparing their success rates on original versus transformed buggy functions.
We prompted each model 10 times per original bug and 10 times per transformed bug, following the recommendations~\cite{arcuri_hitchhikers_2014}, with each sample run in a separate and independent chat session~\cite{sallou_breaking_2024}. Closed-source models were asked to generate 5 patches per prompt; open-source models generated 1 patch per prompt due to context and output limitations. We consider a prompt successful if at least one of the suggested patches passes all tests. This reflects a realistic scenario where unique solutions are generated, even if only one solution exists. 
We define the performance of the model on a given bug in terms of the Success Rate ($SR$): 
\begin{equation}
\small
SR = \frac{\#\text{prompts that produce} \geq 1 \text{ correct patch}}{\#\text{prompts}}
\end{equation}

This is the percentage of prompts that result in at least one plausible patch. This suggests that a prompt is considered successful if at least one of the proposed fixes satisfies all the tests. We can consider this to be a measure that is meaningful in practical situations. So, taking a large number of samples and considering the bug to be solved when at least one patch is correct, like in \cite{xiang_how_2024}, may be acceptable to measure research progress, but it is not feasible in real-world scenarios. We define the success rates on the original and transformed functions as $SR_{orig}$ and $SR_{trans}$. We define the difference in the success rate as $SR_{diff} = SR_{trans} - SR_{orig}$. 


To assess statistical significance between paired results, we use the Wilcoxon signed-rank test. Effect sizes are reported using the Vargha--Delaney $\hat{A}_{12}$ statistic~\cite{vargha2000critique}, which estimates the probability that a randomly selected transformed result outperforms an original one.  A value of $\hat{A}_{12} = 0.5$ indicates no difference; values above 0.5 suggest the transformed version tends to perform better, while values below 0.5 favor the original version. Following standard guidelines~\cite{vargha2000critique}, $\hat{A}_{12}$ values are interpreted as negligible ($0.44 \leq \hat{A}_{12} \leq 0.56$), small ($0.36\!-\!0.44$ or $0.56\!-\!0.64$), medium ($0.29\!-\!0.36$ or $0.64\!-\!0.71$), or large ($\hat{A}_{12} \leq 0.29$ or $\hat{A}_{12} \geq 0.71$). Results are considered statistically significant for $p$-values$<$0.05.

All metrics are based on test-adequate patches, i.e., patches that compile and pass all test cases. While test adequacy is not a guarantee of semantic correctness, manual validation is infeasible at the scale of our study, which includes over 1,000 original and 1,000 transformed bugs evaluated across seven models with multiple repetitions. Test outcomes nevertheless provide a practical and consistent basis for comparing robustness under metamorphic transformations.
\begin{table*}[t]
    \centering
    \caption{Success rate statistics (mean $\pm$ $\sigma$) for bug repairs using original versus semantically transformed code snippets on Defects4J and GitBug-Java. Analysis limited to solvable bugs (original SR $> 0$\%). Statistical significance assessed via Wilcoxon signed-rank test and Vargha-Delaney ($\hat{A}_{12}$) statistics.}
    \label{tab:sota-d4j}
    \scriptsize
    \setlength{\tabcolsep}{3pt}
    \renewcommand{\arraystretch}{1.05}
    \resizebox{\textwidth}{!}{%
\begin{tabular}{llcrcrrcrrcl}
\toprule
\multirow{2}{*}{Dataset} & \multirow{2}{*}{Model} & \multirow{2}{*}{Solvable Bugs} & \multicolumn{2}{c}{Original SR} & \multicolumn{2}{c}{Transformed SR} & \multicolumn{2}{c}{Difference} & \multicolumn{2}{c}{Statistical Tests} \\
\cmidrule(lr){4-5}
\cmidrule(lr){6-7}
\cmidrule(lr){8-9}
\cmidrule(lr){10-11}
 &  &  & Mean & $\sigma$ & Mean & $\sigma$ & Mean & Worst Case & $p$-value & $\hat{A}_{12}$ \\
\midrule
\multirow{7}{*}{Defects4J}
& Claude-3.7    & 409 & 77.31\% & 31.87\% & 72.42\% & 36.06\% & -4.89\%  & -100.00\% & $<$0.001 & 0.414 (Small) \\
& GPT-4o        & 349 & 65.93\% & 34.09\% & 61.92\% & 36.74\% & -4.01\%  & -80.00\%  & $<$0.001 & 0.404 (Small) \\
& GPT-4o-mini   & 279 & 57.17\% & 35.10\% & 51.36\% & 37.77\% & -5.81\%  & -70.00\%  & $<$0.001 & 0.351 (Medium) \\
& Gemma-2-27B   & 201 & 44.38\% & 31.36\% & 36.07\% & 33.87\% & -8.31\%  & -80.00\%  & $<$0.001 & 0.313 (Medium) \\
& Llama-3.1     & 179 & 35.08\% & 26.10\% & 19.11\% & 20.12\% & -15.98\% & -60.00\%  & $<$0.001 & 0.109 (Large) \\
& Ministral-7B  & 110 & 25.64\% & 22.20\% & 18.73\% & 23.97\% & -6.91\%  & -50.00\%  & $<$0.001 & 0.282 (Large) \\
& StarCoder2    & 36  & 13.89\% & 6.45\%  & 6.11\%  & 9.03\%  & -7.78\%  & -30.00\%  & $<$0.001 & 0.208 (Large) \\
\midrule
\multirow{3}{*}{GitBug-Java}
& Claude-3.7    & 37 & 68.11\% & 33.82\% & 68.38\% & 39.05\% & 0.27\%  & -30.00\% & 0.819 & 0.473 (Negligible) \\
& GPT-4o        & 17 & 50.59\% & 37.50\% & 52.35\% & 41.76\% & 1.76\%  & -40.00\% & 0.667 & 0.471 (Negligible) \\
& GPT-4o-mini   & 15 & 55.33\% & 35.23\% & 46.67\% & 34.57\% & -8.67\% & -70.00\% & 0.074 & 0.333 (Medium) \\
\bottomrule
\end{tabular}%
}
\end{table*}

\textbf{Correlation analysis.} To answer \textbf{RQ2}, we investigate whether the performance drop under metamorphic transformation is related to model familiarity, as estimated by the negative log-likelihood (NLL) reported by Ramos et al.~\cite{ramos_are_2024}. These NLL values capture the average uncertainty of each model on the \textit{original} Java file containing the bug and serve as a proxy for potential data leakage. Since Ramos et al. only evaluated a subset of Defects4J, we extend their analysis to the full benchmark using their replication package. It is worth noting that we focus on NLL in this analysis, as Ramos et al.\ found it more sensitive to memorization than 5-gram accuracy, making it a stronger proxy for data leakage.

Because NLL values differ in scale across models, we normalize them using percentile ranks within each model, allowing meaningful comparisons of relative familiarity. We then assess the relationship between the normalized NLL values and the observed performance drop ($SR_{diff}$) using the Spearman rank correlation coefficient, a non-parametric measure that is appropriate given that, as previously determined via the Shapiro–Wilk test, our data do not follow a normal distribution. All correlations are computed using the standard Python implementation of the Spearman test.

\textbf{Analysis of transformation impact.} 
To address \textbf{RQ3}, we analyze which transformation types and combinations most strongly contribute to success rate differences ($SR_{diff}$), serving as indicators of code patterns that are more likely to be memorized by LLMs.
To this aim, we employ a \textit{permutation test}\footnote{\label{fn:anova}We use the \texttt{lmPerm} package from the \href{https://cran.r-project.org/web/packages/lmPerm/index.html}{CRAN repository}.}. This test determines whether the variance in the dependent variable ($SR_{\mathit{diff}}$) can be attributed to the independent variables, which encode the frequency of each transformation type (e.g., the number of times variable names were changed). The permutation test is a non-parametric alternative to two-way Analysis of Variance (ANOVA), and does not rely on assumptions about data distribution. To justify its use, we conducted a Shapiro–Wilk test for normality, which yielded statistically significant $p$-values$<$0.01, indicating that the data deviate from a normal distribution. This violation of the normality assumption further supports our choice of a non-parametric test. We set the number of iterations for the test to 1,000,000 to ensure both stability and reproducibility.

A statistically significant result ($p$-value$<$0.05) suggests that one or more transformations, or their interactions, have a significant effect on the LLM's success rate in generating test-adequate patches. To complement the significance test, we report effect sizes using the Vargha–Delaney $\hat{A}_{12}$ statistic~\cite{vargha2000critique}, which quantifies both the magnitude and direction of the observed effect. We restrict our analysis to interaction terms involving at most three transformation types at a time. This decision is driven by practical concerns. First, higher-order interactions reduce interpretability. Second, the number of transformation combinations grows exponentially, increasing model complexity and the risk of multiple testing errors. Third, estimating higher-order effects requires substantially more data. Combinations involving more than three transformations frequently lead to sparse or empty design cells, which compromises statistical validity and could lead to overfitting.


\section{Results}
\label{sec:results}

\subsection{{RQ1: LLM-based APR Performance Changes}}

\textbf{Results for Defects4J}.
\Cref{tab:sota-d4j} reports the mean and standard deviation of success rates for state-of-the-art APR models on both the original and transformed versions of the bugs in Defects4J and GitBug-Java datasets, averaged over 10 independent runs or prompts. For this analysis, we focus exclusively on bugs that the model was able to solve in the original version (i.e., success rate $>$0\%), which we refer to as \textit{solvable bugs}. This filtering allows us to isolate the impact of semantic transformations on cases where the model has demonstrated bug-fixing capability, allowing us to assess whether such transformations impair or improve performance.

As shown in \Cref{tab:sota-d4j}, all seven models exhibit a statistically significant drop in the success rate when evaluated on the transformed version of Defects4J (all $p$-values are smaller than 0.001). The largest average drop is observed for Llama-3.1 (–15.98\%), with a \textit{large} effect size based on the Vargha-Delaney $\hat{A}_{12}$ statistic. 
Even the smallest observed drop for GPT-4o (–4.01\%) is statistically significant and accompanied by a \textit{small} effect size.
A closer examination of individual bug performance reveals notable worst-case scenarios in which LLMs fail entirely on transformed versions of bugs they previously solved. For example, Claude-3.7 was able to generate a correct patch for \texttt{Jsoup-15} in all 10 runs on the original version, but failed in every run after transformation. GPT-4o exhibited a similar pattern on \texttt{Math-32}, with performance dropping from 80\% to 0\%, as did Gemma-2-27B on \texttt{Lang-52}. Llama-3.1 and Ministral-7B also showed complete failure on \texttt{Cli-11} and \texttt{JacksonDatabind-93}, respectively, despite partial success in the original setting. StarCoder2, although generally less effective, also failed to repair \texttt{Codec-17} after transformation (30\% to 0\%). GPT-4o-mini showed a less extreme but still substantial drop on \texttt{Compress-26}, decreasing from 100\% to 30\%. These results confirm a key robustness vulnerability: even semantically preserving (and natural) code transformations can significantly impair model performance on previously solvable bugs.

To further analyze the results on Defects4J, we examine how performance differences vary with bug difficulty. We categorize bugs into three levels based on the model’s original success rate: \textit{Hard} (0–30\%), \textit{Medium} (30–70\%), and \textit{Easy} (70–100\%), which reflect the empirical difficulty encountered by each model. Hard bugs are rarely solved even in the original form, indicating fundamental challenges; medium bugs are solved inconsistently; and easy bugs are reliably repaired, suggesting they fall well within the model’s capabilities. Metamorphic transformations thus affect bugs differently depending on their empirical difficulty.

\Cref{fig:sota-d4j} depicts the change in success rates for each model across the three difficulty levels. Negative values (below the zero line) indicate a performance drop after transformation, while positive values (above the line) indicate an improvement. Across all models, we observe a consistent drop in performance for each difficulty level, with the most pronounced declines occurring on \textit{easy} bugs. 

For \textit{hard} bugs, where models already struggle to produce correct patches, the median performance remains close to zero both before and after transformation. In these cases, metamorphic transformations have limited additional impact, unsurprisingly, as the model fails even without transformation. Nonetheless, we observe a few positive outliers, suggesting that in rare instances, transformations may make certain hard bugs more tractable, potentially by simplifying the syntax or nudging the model toward a more effective solution. This observation opens an interesting avenue for future work: systematically identifying metamorphic transformations that can assist LLMs in overcoming failure cases, rather than merely serving as robustness checks.

\begin{figure}
    \centering
    \includegraphics[width=0.9\linewidth]{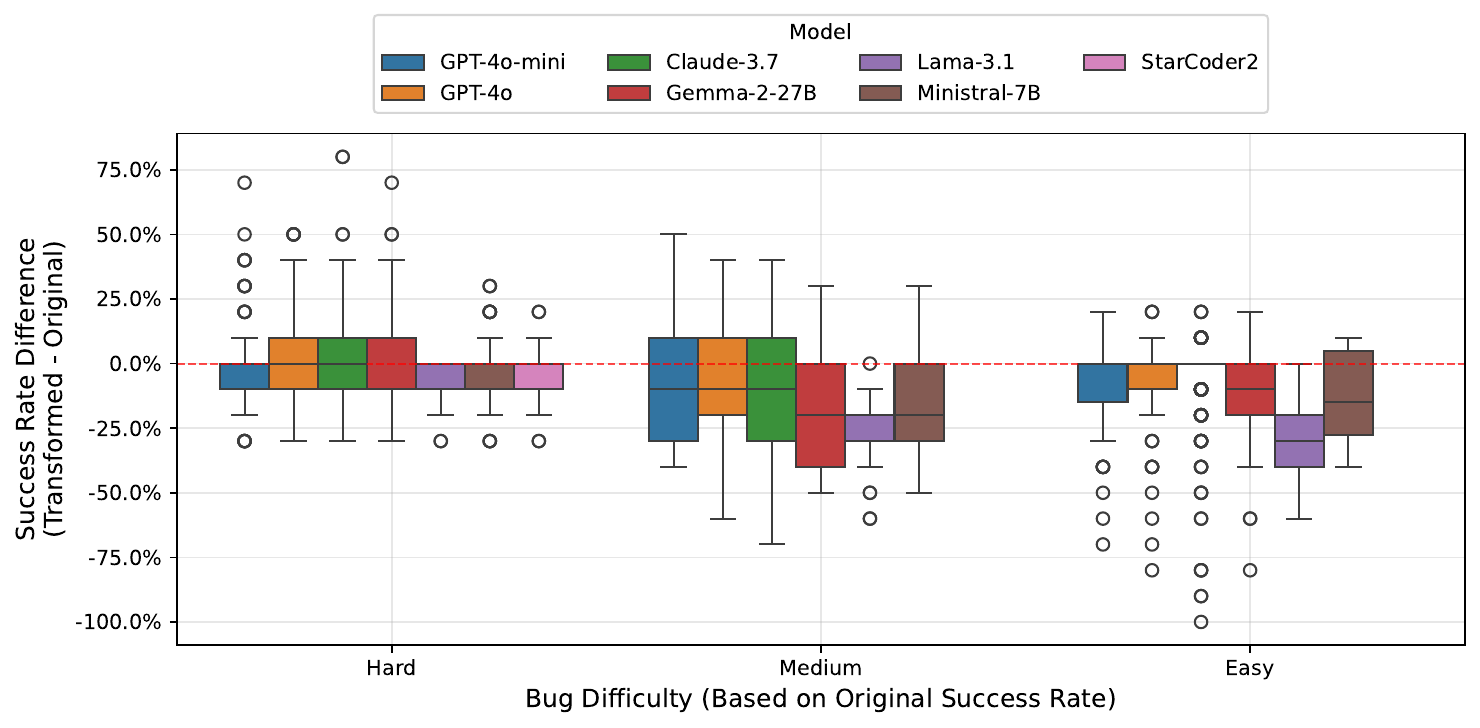}
    \caption{Distribution of success rate differences ($SR_{diff}$) for each model after metamorphic transformations, grouped by bug difficulty levels on the Defects4J benchmark. 
    }
\label{fig:sota-d4j}
    \label{fig:sota-d4j}
\end{figure}

For \textit{medium} bugs, all models exhibit a clear median drop of approximately –10\%, indicating that transformations disrupt model behavior in cases where success is already inconsistent. For \textit{easy} bugs, the effect is most severe: all models show a decline in median, 25th, and 75th percentile performance, with Llama-3.1 and Mistral-7B showing median drops exceeding –20\%. These results indicate that even for seemingly trivial repair tasks, metamorphic transformations can substantially hinder model performance —especially for weaker models— highlighting the fragility of LLM-based APR.

\textbf{Results for GitBug-Java}.
We focus on Claude-3.7-Sonnet, GPT-4o, and GPT-4o-mini. These models have (pre)train\-ing cutoffs in October 2023\footnote{\url{https://learn.microsoft.com/en-us/azure/ai-foundry/openai/concepts/models}} (GPT-4o and GPT-4o-mini) and approximately November 2023\footnote{\url{https://docs.anthropic.com/en/docs/about-claude/models/overview}} (Claude-3.7), which closely match the commit window of GitBug-Java that spans from January to October 2023~\cite{silva_gitbug-java_2024}. This temporal alignment enables the investigation of potential memorization effects, as any leakage would likely stem from pretraining exposure.

\begin{figure}
    \centering
    \includegraphics[width=0.9\linewidth]{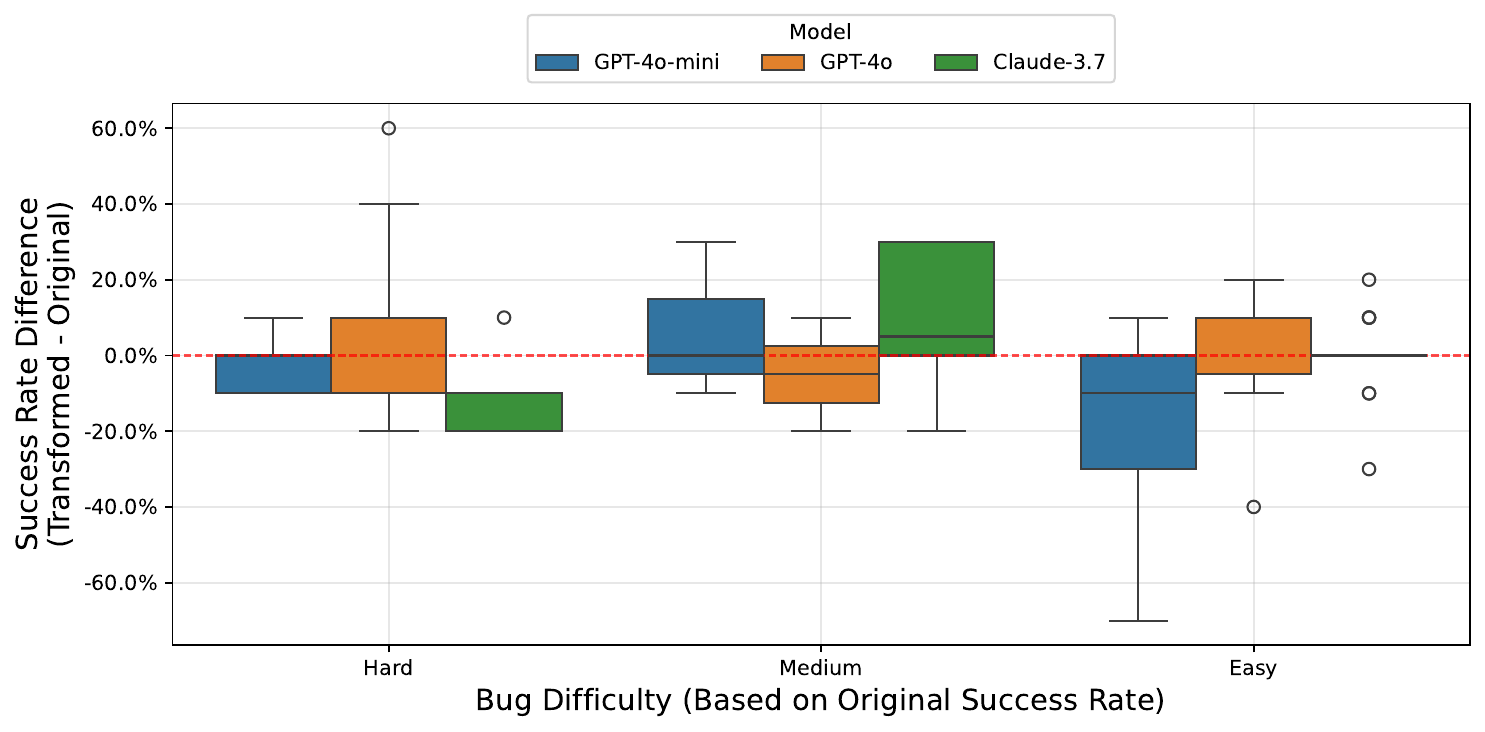}
    \caption{Distribution of success rate differences ($SR_{diff}$) for each model after metamorphic transformations, grouped by bug difficulty levels on the GitBug-Java benchmark. 
    }
    \label{fig:sota-gitbug-difficulty}
\end{figure}

The results for this benchmark are shown in \Cref{tab:sota-d4j}. Unlike Defects4J, we do not observe consistent performance degradation across models. For Claude-3.7-Sonnet and GPT-4o, success rates on transformed bugs are slightly higher than on the original versions, with negligible effect sizes and no statistical significance ($p = 0.819$ and $p = 0.667$, respectively). GPT-4o-mini shows a small performance drop of 8.67\%, which is marginally significant ($p = 0.074$) with a medium effect size ($\hat{A}_{12} = 0.333$). Despite the absence of broad performance drops, all three models experience notable worst-case degradations. For example, GPT-4o-mini drops from 100\% to 30\% on \path{nikoo28-java-solutions-8d81307ea165}, GPT-4o drops from 90\% to 50\% on \path{traccar-traccar}-\texttt{6f59f756a7d3}, and Claude-3.7 drops from 100\% to 70\% on \path{traccar-traccar-5c26f25b3b0a}. These cases show that even when average performance remains stable, certain transformed bugs become substantially more difficult for models that previously succeeded.

We note that GitBug-Java is a more challenging benchmark overall. Most bugs in the dataset cannot be fixed by any model in our study, and the number of \textit{solvable bugs} (i.e., those with a non-zero success rate) is substantially smaller than in Defects4J. As shown in \Cref{fig:sota-gitbug-difficulty}, this limits the statistical power of per-difficulty analysis and may partially explain the absence of clear trends. Nonetheless, the observed variation and worst-case drops suggest that metamorphic transformations can still affect performance, although less uniformly than in older benchmarks such as Defects4J.

\begin{tcolorbox}[colback=gray!10, colframe=black, boxrule=0.5pt, sharp corners]
\textbf{Answer to RQ1:} State-of-the-art LLMs perform significantly worse on the metamorphically transformed Defects4J benchmark. They do not suffer as much on the GitBug-Java Benchmark. Certain bugs reveal model vulnerabilities to metamorphic transformations.
\end{tcolorbox}

\subsection{{RQ2: Performance drop and data leakage}} 

\begin{figure}
    \centering
    \includegraphics[width=0.95\linewidth]{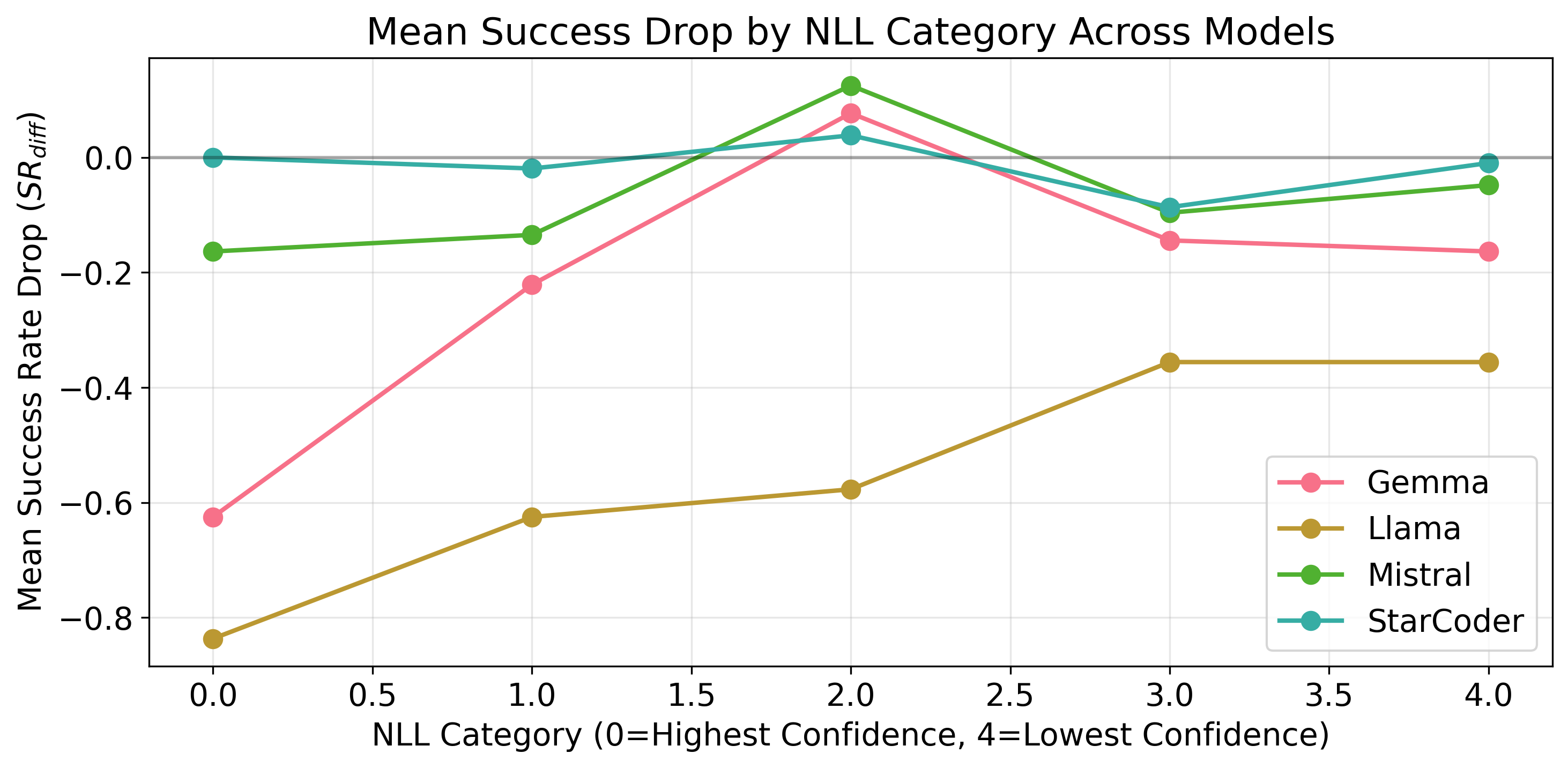}
    \caption{Mean performance drop ($SR_{diff}$) across NLL confidence categories. Bugs are grouped into five categories based on their NLL percentile ranks (from lowest to highest NLL). Lower categories indicate higher model confidence (and likely familiarity), while higher categories indicate lower confidence.
    }
    \label{fig:nll-categorical-drop}
\end{figure}

\Cref{fig:nll-categorical-drop} shows the mean success rate drop ($SR_{diff}$) across NLL-based confidence categories. Bugs are grouped into five bins according to their NLL percentile, where category 0 includes the most familiar (lowest NLL) and category 4 the least familiar (highest NLL) instances. Notice that for this analysis, we focus on the four open-source models: Gemma, Llama, Mistral, and StarCoder. NLL values are not available for proprietary models because computing NLL requires access to token-level probabilities from the model’s forward pass, which in turn requires access to the model’s internal weights and tokenizer. Since this is restricted for closed-source models like GPT-4 or Claude-3.7, correlation analysis involving NLL is not feasible for these models.


From \Cref{fig:nll-categorical-drop}, we see a clear trend: for all models, the average performance drop is highest in the lowest NLL bin. This suggests that metamorphic transformations disproportionately impact bugs the model is more familiar with, as shown by low NLL values. The consistent link between performance degradation and NLL offers further evidence that combining metamorphic transformations with model familiarity signals is effective for revealing data leakage. Interestingly, we also observe that performance drop increases again in the highest NLL bins for Gemma-2, Mistral-7B, and StarCoder2, forming an inverted U-shape. This suggests a second source of vulnerability: for unfamiliar code, the models are highly uncertain and sensitive to perturbations, leading to prediction instability. Metamorphic testing serves two complementary purposes: (1) \textit{diagnose data leakage (together with low NLL) by disrupting memorized inputs}, and (2) \textit{reveal brittleness in model predictions under uncertainty}.

To statistically assess the relationship between model familiarity and robustness, we compute the Spearman rank correlation between NLL and $SR_{diff}$ for the lower half of the NLL distribution. For Gemma-2, we observe a correlation of $r$=0.303 ($p$-value=0.0011), indicating a \textit{medium} effect size. Llama-3.1 and Mistral-7B also show positive correlations of $r$ = 0.165 ($p$-value=0.0472) and $r$=0.218 ($p$-value= 0.0491), respectively, both with \textit{small} effect sizes. StarCoder2 shows a stronger correlation of $r$=0.401 ($p$-value=0.0798), which falls just outside the conventional significance threshold due to limited data, but still indicates a \textit{medium} effect size. These findings confirm that performance degradation is associated with model familiarity, reinforcing metamorphic transformations as a diagnostic and decontamination LLM-based APR.

\begin{figure}
    \centering
    \includegraphics[width=0.85\linewidth]{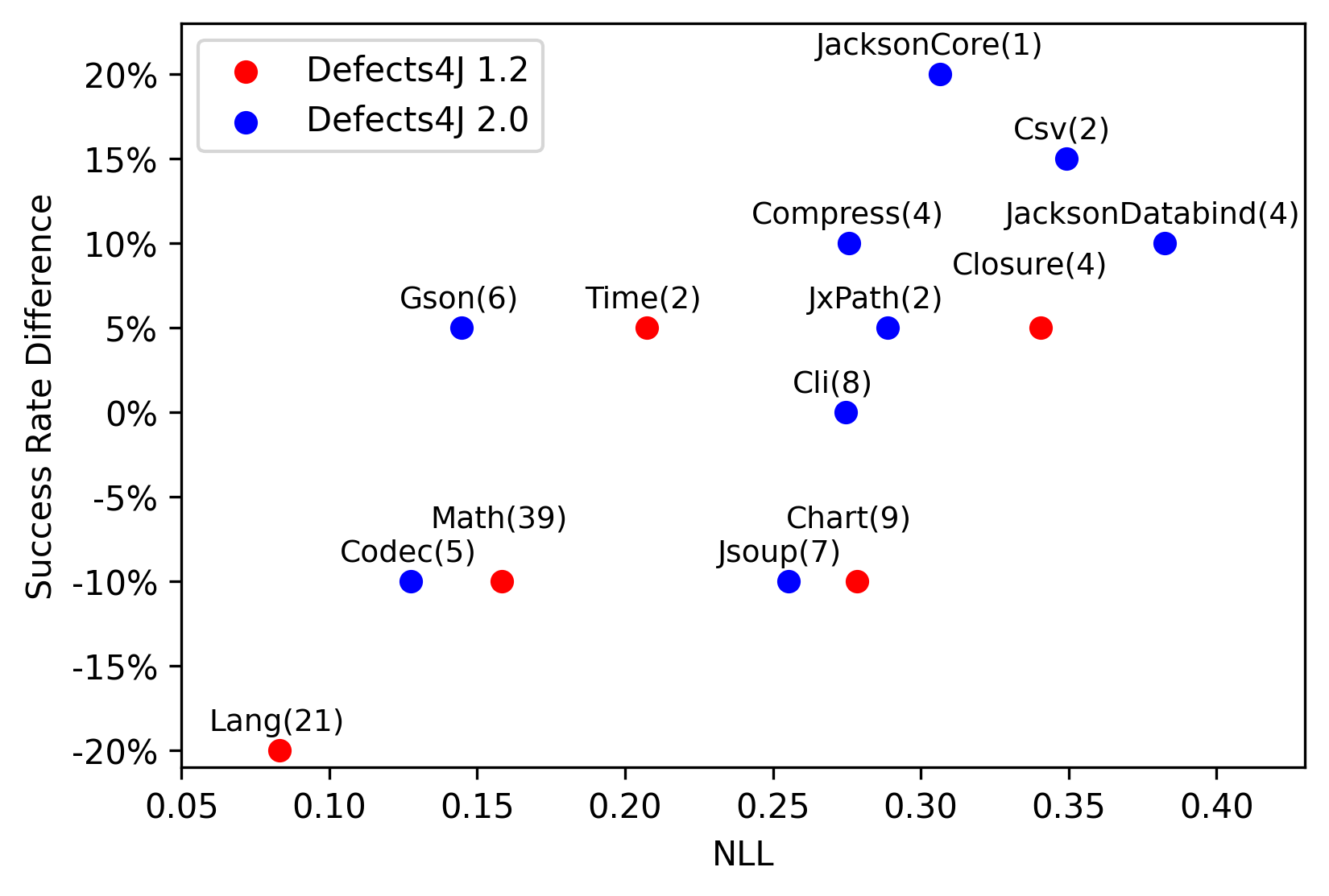}
    \caption{Defects4J projects by average success rate drop, median NLL, and Defects4J version. Only bugs solvable by Gemma 2 27B with sub-median NLL are shown. Bug counts per project are noted in parentheses.}
    \label{fig:gemma_scatter}
\end{figure}

Further inspection of Gemma-2, the model with the strongest correlation between NLL and performance drop, reveals that many low-NLL bugs originate from the \verb|Lang| and \verb|Math| projects. As shown in \Cref{fig:gemma_scatter}, these projects exhibit both low median NLL and large success rate drops. Similar patterns are observed for \verb|Codec|. These trends suggest that Gemma-2 is familiar with these codebases and that its APR performance may be inflated by data leakage. Furthermore, \Cref{fig:gemma_scatter} shows that projects from Defects4J version 1.2 tend to have lower NLL and higher performance drops, reinforcing the idea that the benchmark age and prior exposure during pretraining contribute to memorization effects.



\begin{tcolorbox}[colback=gray!10, colframe=black, boxrule=0.5pt, sharp corners]
\textbf{Answer to RQ2:} 
We find a significant correlation between model familiarity (low NLL) and success rate drops under metamorphic transformations for all models (only marginal for StarCoder2 due to limited data). 
Metamorphic transformations and model familiarity signals offer complementary evidence of data leakage.\end{tcolorbox}

\subsection{{RQ3: Transformations indicative of memorization}}

\Cref{tab:transformations} summarizes statistically significant associations between individual or combined transformations and success rate differences, identified using a two-way permutation test ($p$-value$<$0.05). Effect sizes are reported using the Vargha–Delaney $\hat{A}_{12}$ statistic, where $\hat{A}_{12}<$ 0.5 indicates that the transformation(s) are associated with lower repair success. A negative $SR_{diff}$ indicates that models repair fewer bugs with transformed code than vs. the original version.


Across models, the transformation most consistently associated with performance degradation is \textit{NestElseIf}, which is significant for both Claude-3.7 and GPT-4o, both in isolation and in combination with other transformations. This transformation replaces an \texttt{else-if} chain with nested \texttt{if} statements (see \Cref{tab:chosen_transformations}), increasing structural complexity while preserving semantics. Such sensitivity suggests that models may rely on familiar syntactic patterns. 

For Claude-3.7, \textit{NestElseIf} exhibits stronger effects when combined with other transformations, including \textit{ExpandUnaryIncrement}, \textit{ForToWhile}, \textit{RenameFunction}, and \textit{RenameParameters}. This pattern indicates increased difficulty when structural changes are compounded with lexical variation, consistent with reliance on memorized code forms.

GPT-4o and GPT-4o-mini show fewer significant associations overall. For GPT-4o, \textit{NestElseIf} alone yields a small but significant effect, while both models exhibit performance drops when parameter renaming is combined with other transformations such as \textit{ExpandUnaryIncrement} or \textit{NestElseIf}. This suggests that renaming acts as an amplifying factor when paired with structural perturbations. 

Renaming transformations (\textit{RenameVariables} and \textit{RenameParameters}) generally have a significant adverse effect only when combined with structural changes, indicating that renaming alone rarely impairs performance but can exacerbate degradation when paired with other transformations. Gemma-2-27B is a notable exception: it is negatively affected by renaming in isolation and also shows medium-sized effects for \textit{SwapEqualsOperands}, with effects becoming large when equality and relational operand swaps are combined. This behavior indicates a stronger reliance on surface-level variable names and expression structure, making the model particularly sensitive to deviations from familiar code patterns.

Mistral-7B also exhibits a significant drop in performance when the \textit{ForToWhile} transformation is applied, indicating that it performs better when loops are expressed using \texttt{for} rather than \texttt{while}. The model also shows a performance drop when \textit{RenameParameters} is combined with \textit{SwapRelationOperands}.

Finally, no specific transformation is identified as disproportionately harmful for Llama-3.1 or StarCoder2. Llama-3.1 exhibits broadly degraded performance across nearly all transformations, limiting the ability of the permutation test to isolate individual effects. In contrast, StarCoder2 repairs very few bugs overall (36 in total). This low number of bug fixes results in very few data points for each transformation, limiting the statistical power of the permutation test. In other words, the sample size was too small to detect significant interactions between transformations and performance drops reliably.


\begin{tcolorbox}[colback=gray!10, colframe=black, boxrule=0.5pt, sharp corners]
\textbf{Answer to RQ3:} Certain metamorphic transformations are consistently associated with larger performance drops, indicating models sensitivity to these code patterns. \textit{NestElseIf} is the most indicative transformation across models, particularly for Claude-3.7 and GPT-4o. Renaming lead to performance degradation when combined with structural changes, while Gemma-2-27B is uniquely sensitive to renaming in isolation.
\end{tcolorbox}

\section{Discussion and Implication}
\label{sec:discussion}




\begin{table}[t]
\centering
\caption{Statistical interactions between individual or combined transformations and the drop in success rate based on the two-way permutation test ($p$-value$<$0.05). The effect size of $\hat{A}_{12}$$<$0.5 indicate a harmful impact on the APR success rate.
}
\label{tab:transformations}
\resizebox{\linewidth}{!}{
\begin{tabular}{llcc}
\toprule
\textbf{Model} & \textbf{Transformations} & \textbf{p-value} & \textbf{$\hat{A}_{12}$} \\
\midrule

\multirow{6}{*}{Claude-3.7}
& NestElseIf & $<$0.01 & 0.41 (small)) \\
& ExpandUnaryIncrement + NestElseIf & $<$0.01 & 0.41 (small) \\
& ForToWhile + NestElseIf & $<$0.01 & 0.41 (small) \\
& NestElseIf + RenameFunc. & $<$0.01 & 0.41 (small) \\
& NestElseIf + RenameParam. & $<$0.01 & 0.41 (small) \\
& NestElseIf + SwapRelationOper. & $<$0.01 & 0.37 (small) \\
\cmidrule{2-4}

\multirow{2}{*}{GPT-4o}
& NestElseIf & $<$0.01 & 0.44 (small) \\
& NestElseIf + RenameParameters & $<$0.01 & 0.44 (small) \\
\cmidrule{2-4}

\multirow{1}{*}{GPT-4o-mini}
& ExpandUnaryIncrement+RenameParameters & $<$0.02 & 0.42 (small) \\
\cmidrule{2-4}

\multirow{5}{*}{Gemma-2-27B}
& RenameVariables & 0.03 & 0.31 (medium) \\
& SwapEqualsOperands & 0.01 & 0.32 (medium) \\
& ExpandUnaryIncrement + ReverseIf & 0.03 & 0.32 (medium) \\
& SwapEqualsOperands + SwapRelationOperands & 0.04 & 0.23 (large)\\

& RenameParameters + RenameVariables & 0.04 & 0.31 (medium) \\ 
& ReverseIf + SwapEqualsOperands & 0.04 & 0.32 (medium) \\
\cmidrule{2-4}


\multirow{1}{*}{Mistral-7B}
& ForToWhile & 0.04 & 0.39 (small) \\
& RenameParameters + SwapRelationOperands & 0.02 & 0.41 (small) \\
\bottomrule
\end{tabular}
}
\end{table}

\textbf{Scope and Positioning}. This work is not a proposal for a new automated program repair technique, nor a replacement for benchmark curation or post-cutoff dataset construction. Instead, it addresses a complementary evaluation problem: assessing whether reported APR performance reflects genuine reasoning or is inflated by memorization. We position metamorphic testing as a lightweight, model-agnostic diagnostic that can be applied selectively to stress-test evaluation results. Our goal is not to eliminate data leakage at training time, but to reduce the inflation of reported performance and improve the interpretability and robustness of empirical APR evaluations in the LLM era. Throughout this paper, we use the term \emph{mitigation} strictly in this evaluation sense, rather than to denote preventing or eliminating data leakage during model training.
While prior work has independently studied robustness under MT and benchmark memorization via model-internal metrics, our study provides empirical evidence linking these two perspectives through observable repair behavior in LLM-based program repair.

\textbf{Understanding the Fragility of LLM-Based APR}.
Our results show that even state-of-the-art LLM-based automated program repair (APR) systems are sensitive to natural, semantics-preserving code transformations. On Defects4J, all evaluated models exhibit statistically significant performance degradation, including on bugs that were previously easy to fix. Thus, strong benchmark performance does not necessarily imply robust reasoning, but may partially rely on familiarity with recurring syntactic patterns. In contrast, performance on the more recent GitBug-Java benchmark remains largely stable on average, though notable worst-case degradations persist. These findings highlight that average-case metrics can mask brittleness at the instance level, and that robustness failures may still arise even on leakage-aware benchmarks.

\textbf{Metamorphic Testing as a Diagnostic Tool}.
Performance drops under MT are especially pronounced for instances that models seem most familiar with, as reflected by low NLL values. Although NLL by itself is not proof of data leakage, the consistent link between low NLL and greater performance degradation gives further evidence that memorization may inflate reported APR performance.
Importantly, we position metamorphic testing not as a definitive leakage detector, but as a \emph{model-agnostic diagnostic probe}. Unlike metrics that rely on access to model internals, metamorphic testing exposes code-pattern sensitivity through observable behavior. This makes it applicable to both open-source and proprietary models and suitable for practical evaluation settings.

\textbf{What Transformations Reveal About Memorization}.
Our RQ3 analysis shows that certain transformations are more indicative of memorization effects than others. Structural rewrites such as \textit{NestElseIf} consistently expose performance drops across multiple models, particularly when combined with lexical variation. Renaming transformations alone rarely impair performance, but amplify degradation when paired with structural changes. Gemma-2-27B stands out as uniquely sensitive to renaming in isolation, suggesting a stronger reliance on surface-level identifiers and expression structure. These patterns reinforce the view that LLMs may rely on familiar syntactic forms rather than abstract program semantics.

\textbf{Dataset Age and Evaluation Validity}.
Differences between Defects4J and GitBug-Java highlight the importance of dataset design. Older benchmarks, such as Defects4J, are more likely to overlap with pretraining data and exhibit stronger memorization signals, whereas newer benchmarks reduce but do not eliminate these effects. Our results suggest that even leakage-aware datasets can benefit from robustness-oriented evaluation, as memorization traces may still emerge at the instance level.

\textbf{Looking Beyond Averages}.
Although mean success rates on GitBug-Java show limited change under transformation, individual bugs can become substantially harder or unsolvable. This underscores the limitations of relying solely on aggregate metrics and motivates the use of robustness analyses to complement standard APR evaluations.



\textbf{Implications for Future Evaluations}.
We do not argue that MT replaces benchmark curation or post-cutoff dataset construction. Rather, it offers a complementary, low-overhead mechanism to assess whether reported performance is robust to natural variation. Transformations can be applied selectively (for example, to likely memorized instances), making the approach practical without imposing prohibitive evaluation costs. Reporting results on both original and transformed benchmarks can therefore help reduce the risk of performance inflation and improve the interpretability of APR evaluations. Our results further suggest that metamorphic testing need not be applied exhaustively: focusing on a small subset of instances with strong memorization signals (e.g., the lowest 20--25\% by NLL) already exposes substantial performance inflation, keeping evaluation overhead modest.

\textbf{Comparison with Prior Work}. 
Our findings confirm prior evidence that code models are fragile under metamorphic transformations~\cite{applis_assessing_2021, applis_searching_2023, yang_natural_2022}, and extend these observations to state-of-the-art LLM-based APR systems. Unlike earlier studies that focused primarily on robustness, we explicitly link transformation sensitivity to memorization signals, corroborating and extending recent leakage analyses~\cite{ramos_are_2024}. In doing so, we provide large-scale empirical evidence, previously missing in the literature, that metamorphic transformations,  when considered alongside model familiarity signal, can serve as a practical, model-agnostic diagnostic for memorization effects in LLM-based program repair.

\section{Threats to Validity}

\textbf{Threats to internal validity.} %
First, the \texttt{Codecocoon} implementation may generate incorrect or unnatural variants. We mitigate this risk through extensive unit testing and manual inspection. Second, bugs in the experimental pipeline could affect results; however, such issues would impact original and transformed variants equally (except in the transformation stage), which we manually validated. Third, updating function names in tests and stack traces relies on regular expressions that may miss some references. We manually verified that unmatched occurrences did not refer to the target function (e.g., substrings in variable/test names).
Metamorphic transformations may still produce code that appears less natural, potentially disadvantaging APR models trained on human-written code and contributing to observed performance drops. To mitigate this, we restrict our study to natural transformations~\cite{yang_natural_2022, le-cong_evaluating_2024} and use an LLM to generate semantically appropriate identifier synonyms. While NLL is correlated with code naturalness~\cite{ramos_are_2024}, it is computed on the \emph{original} code and is unaffected by metamorphic transformations, limiting its impact as a confounding factor.

\textbf{Threats to external validity.}
Since the dataset provided by Ramos et al.~\cite{ramos_are_2024} covers only a subset of Defects4J (\verb|Closure|, \verb|Lang|, \verb|Chart|, \verb|Math|, and \verb|Mockito|), their benchmark may not fully represent all subject in Defects4J. We reproduced their method in the remaining projects to collect more samples. 
While the APR community is increasingly exploring newer benchmarks, Defects4J remains the most widely used evaluation dataset in recent APR research~\cite{yang2025surveyllmbasedautomatedprogram}. Understanding evaluation bias in such dominant benchmarks is therefore critical, as conclusions drawn from them continue to shape research directions.
Our results are limited to APR. While we expect metamorphic transformations to have the same impact on other tasks (e.g., code or test generation) this generalization remains to be confirmed.


\textbf{Threats to construct validity.} We have used the frequency of patches that pass all tests as the performance metric to evaluate APR performance. However, a patch may pass the suite but fail to fully fix the bug~\cite{zhang_survey_2023} (\textit{test overfitting}). Other studies use exact-match metrics~\cite{huang_survey_2023, chen_neural_2023, chen_sequencer_2021} or manual correctness checking~\cite{xiang_how_2024, xia_automated_2024}. We chose test-suite pass rate since exact-match metrics penalize bugs with multiple valid solutions~\cite{huang_survey_2023}, and manual check is infeasible. Although imperfect, this metric is applied equally to original and transformed groups, making it unlikely that overfitting biases one group more. 
We applied each transformation wherever possible in code. So, a transformation might have been applied multiple times in a larger code snippet.
Hence, we compute not only the correlation between performance drop and transformations, but also between performance drop and bug size/complexity. For example, if a variable name transformation applies once to bug~\textit{A} but three times to bug~\textit{B}, this reflects that bug~\textit{B} has more local variables. Thus, our tests may reflect the impact of 'transformable' features.

\section{Conclusion and Future Work}

We explored metamorphic transformations as an evaluation-time diagnostic strategy for understanding the impact of data leakage. To this end, we applied semantic-preserving transformations to the Defects4J and GitBug-Java benchmarks and demonstrated that state-of-the-art LLMs exhibit noticeable performance degradation (w.r.t. their ability to generate bug fixes) on transformed bugs, particularly on Defects4J. 

We found that metamorphic transformations introducing nested conditionals and those involving identifier renaming were the most harmful overall. Identifier renaming, in particular, significantly degraded the APR performance for Claude-3.7, likely because identifier names act as cues for memorized solutions. We also observed statistically significant (small and medium) correlations between performance drop and negative log-likelihood for Gemma-2, Llama-3.1, and Mistral-7B, supporting the link between model familiarity and degradation in robustness under transformation. Together, metamorphic transformations and model familiarity (low NLL values) provide complementary evidence of data leakage.


In future work, we aim to explore adversarial MT, where transformation combinations are selected to maximally reduce model performance on individual bugs. We also intend to expand our analysis to other code intelligence tasks, such as test case generation, bug localization, and code summarization, as well as to additional benchmarks and programming languages, including Python, C++, Kotlin, and JavaScript.

\bibliographystyle{IEEEtran}
\bibliography{references}
\balance

\end{document}